\documentclass[twocolumn,aps,amsmath,amssymb,amsfonts,showpacs,prd,10pt]{revtex4}
\usepackage{epsfig}
\usepackage{amssymb}
\usepackage{subfigure}
\begin{document}
\title{A simple model for the dynamical Casimir effect for a static mirror with time-dependent properties}
\author{Hector O. Silva}
\email{hokada@ufpa.br}
\affiliation{Faculdade de F\'\i sica, Universidade Federal do
Par\'a, 66075-110, Bel\'em, PA,  Brazil}

\author{C. Farina}
\email{farina@if.ufrj.br}
\affiliation{Instituto de F\'\i sica, Universidade Federal do Rio de Janeiro,
Caixa Postal 68528, 21941-972, Rio de Janeiro, RJ, Brazil}

\date{\today}

\begin{abstract}

We consider a real massless scalar field in 1+1 dimensions
satisfying time-dependent Robin boundary condition at a static
mirror. This condition can simulate moving reflecting mirrors whose
motions are determined by the time-dependence of the Robin
parameter. We show that particles can be created from 
vacuum, characterizing in this way a dynamical Casimir effect. 
\end{abstract}
\maketitle

\section{Introduction}
\label{Intro}

The phenomenon of particle creation from quantum vacuum by moving
boundaries or due to time-dependent properties of materials,
commonly referred to as the dynamical Casimir effect (DCE)
\cite{Yablo-1989,Schwinger-1992}, has been investigated since the
pioneering works of Moore \cite{Moore-JMP-1970} and DeWitt \cite{DeWitt-1975} (see also the subsequent works carried in Refs. \cite{others}) in a wide variety of
situations and with the aid of quite different approaches (see Refs.
\cite{Reviews-I} for excellent reviews on the subject). Particularly, perturbative and numerical approaches
were applied for single mirrors
\cite{Single-mirror,Jaekel-PRL-1996,Mintz-JPA-2006,Mintz-JPA-2006-2} and cavities
\cite{Cavities}. Initial field states different from vacuum were
also considered for single mirrors \cite{temperatura-uma-fronteira,equivalencia-DN}
and cavities as well \cite{temperatura-cavidade}. The first experimental observation of this phenomenon was recently announced in Ref. \cite{Wilson-arXiv}.


Taking into account the difficulties in generating appreciable mechanical oscillation frequencies (of the
order of GHz) to obtain a detectable number of photons, recent
experimental schemes focus on simulating moving boundaries by
considering material bodies with time-dependent electromagnetic
properties. These possibilities were first proposed by Yablonovitch
\cite{Yablo-1989} and have been further developed in
theoretical works which considered materials with time-dependent
permittivities and time-dependent surface conductivities
\cite{time-dependent-prop,Crocce-PRA-2004,
Naylor-PRA-2009} (see the
nice compilation done in Ref. \cite{Naylor-PRA-2009}). For instance,
in Ref. {\cite{Crocce-PRA-2004} the DCE for a massless scalar field
within a cavity containing a thin semiconducting film with
time-dependent conductivity and centered at the middle of the cavity
was studied. The coupling of such a film to the quantum scalar field
was modeled by a delta-potential with time-dependent strength. A
generalization to the case of an electromagnetic field was carried
out in \cite{Naylor-PRA-2009}. Very promising and ingenious
experimental set-ups to simulate non-stationary boundaries include
the changing of the reflectivity of a semiconductor by the incidence
of a periodic sequence of short laser pulses \cite{Experimento-MIR}
or by using a coplanar waveguide terminated by a superconducting
quantum interference device (SQUID). Applying a variable magnetic
flux on the SQUID, a single moving mirror can be simulated
\cite{Johansson,Nation}. A first step toward the experimental verification of the DCE was 
recently made in \cite{Wilson-PRL-2010} using this approach. Moreover, the same group recently claimed to have observed the DCE \cite{Wilson-arXiv}.

Key ingredients in the predictions of the DCE are the boundary
conditions (BC)  under consideration and naturally the quantum field
submitted to those BC. Quite general BC are the so called Robin ones
which, for the case of a scalar field in 1+1 dimensions and a single
mirror fixed at $x=a$, are defined by
$\phi(t,x=a)=\gamma[\partial_{x}\phi(t,x)]_{x=a}$, where $\gamma$ is
a real parameter (called hereafter as Robin parameter). For the case
of a moving boundary, the previous relation is imposed
in the comoving frame and the corresponding BC in the laboratory frame
is obtained after an appropriate Lorentz transformation.

This BC have the nice feature of
interpolating continuously Dirichlet ($\gamma \rightarrow 0$) and
Neumann ($\gamma \rightarrow \infty$) ones and occurs in several
areas of physics and mathematics.
 For instance, in classical mechanics they will appear if one considers a vibrating string coupled
to a spring which satisfies Hooke's law and is localized at one of
its edges \cite{Mintz-JPA-2006,Mintz-JPA-2006-2,Robin-CM}. In
non-relativistic quantum mechanics, Robin BC occur as the most
general  BC imposed by a wall ensuring the hermiticity of the
hamiltonian as well as a null probability flux through it
\cite{Robin-QM}. Regarding the static Casimir effect \cite{Casimir-1948}, it
was shown that the Casimir force between two parallel plates which
impose Robin BC on a real scalar field may have its sign changed if
appropriate choices are made for the corresponding Robin parameters
of each mirror \cite{Romeo-JPA-2002}. Such kind of repulsive Casimir force was also predicted, in the case of parallel plates, by Boyer in the 70s, who considered a pair of perfectly conducting and infinitely permeable plates \cite{Boyer-PRA-1974}. Further investigations on the influence of Robin BC in the static Casimir effect, including thermal corrections and the case of Casimir piston setups,
were carried, for instance, in Refs. \cite{Robin-SCE}. See also Refs. \cite{Quantum-Vacuum} for the influence of this BC on the structure of quantum vacuum.

Only recently Robin BC were considered in the context of the DCE.
For a massless scalar field in 1+1 dimensions submitted to a Robin
BC at a single moving mirror the radiation reaction force on the
moving mirror and the particle creation rate were computed in Refs.
\cite{Mintz-JPA-2006,Mintz-JPA-2006-2}. Interestingly, for Robin BC,
the radiation reaction force acquires a dispersive component, in
sharp contrast with Dirichlet and Neumann cases where the force is
purely dissipative. It was also shown that, for a given Robin
parameter, there exists a mechanical frequency of motion that
dramatically reduces the particle creation effect
\cite{Mintz-JPA-2006-2}. Finally, and of crucial importance for the
present work, Robin BC can also be useful
 to describe phenomenological models for penetrable surfaces and under
certain conditions they simulate the plasma model for real metals
\cite{Robin-EM}. In these situations, for frequencies $\omega$ much
smaller than the plasma frequency
$\omega_{\mbox{\footnotesize{P}}}$, the Robin parameter $\gamma$ can
be identified as the plasma wavelength
$\lambda_{\mbox{\footnotesize{P}}}$.  In other words, the Robin parameter
$\gamma$ gives us an estimative of the penetration length of the
mirror under consideration \footnote{The set of references
concerning the physical applications of Robin BC provided in the
present paper obviously does not intend to be complete. Our objective
is just to give the reader a taste of the richness of physical situations involving this BC.}.

Since to simulate a motion of a reflecting mirror is equivalent to
simulate a real metal with time-dependent plasma wavelength, the
above interpretation of $\gamma$ leads naturally to the
consideration of time-dependent Robin parameters. Specifically
speaking, it is quite natural to simulate the motion of a reflecting
mirror by considering the quantum field submitted to a Robin BC at a
static mirror but with a time-dependent Robin parameter $\gamma(t)$.
 The kind of boundary motion which is being simulated is determined
by the kind of time-dependence of $\gamma(t)$. The purpose of this
paper is precisely to analyze this situation for a massless scalar
field in 1+1 dimensions. Particularly, we shall compute explicitly
the particle creation rate for a natural choice of
time-dependence for $\gamma(t)$ which is directly related to recent
experimental proposals. This paper in organized as follows: in Sec.
\ref{Bogoliubov} the Bogoliubov transformation between the in and
out creation/annihilation operators are obtained, allowing us to
find the spectral distribution of the created particles and the particle creation rate in Sec.
\ref{Spectrum} and \ref{creation_rate}, respectively. Finally, in Sec. \ref{Conc} we present our
conclusions and final remarks. Throughout this work we consider
$\hbar=c=1$.

\section{The Bogoliubov transformation}
\label{Bogoliubov}
We start considering a real massless scalar field $\phi$ in 1+1 dimensions which
satisfies the Klein-Gordon equation, 
$\partial^2\phi = 0$, 
and is
submitted to a time-dependent Robin BC at a mirror fixed at the
origin, namely, 
$\gamma(t)\partial\phi/\partial x|_{x=0} - \phi(0,t)
= 0$. 
For simplicity, we assume that $\gamma(t)$ departs only
slightly from a positive constant $\gamma_0$, so that we can write
$\mbox{$\gamma(t)=\gamma_{0}+\delta\gamma(t)$}$, where
$\delta\gamma(t)$ is a smooth time-dependent function satisfying the
condition $\max|\delta\gamma(t)| \ll \gamma_{0}$, for every $t$.
Under these assumptions in the limit $\gamma_{0} \rightarrow \infty$ we recover
Neumann BC. On the other hand, to reobtain Dirichlet BC ($\gamma_{0} \rightarrow 0$), because of condition $\max |\delta\gamma(t)| \ll \gamma_{0}$, we must also take $\delta\gamma(t) \rightarrow 0$. If we consider only $\delta\gamma(t)=0$ we re-obtain the usual time-independent Robin BC.
Moreover, we shall also
impose that $\delta\gamma(t) \rightarrow 0$ for $t \rightarrow \pm
\infty$. The BC satisfied by $\delta\gamma(t)$ then reads
\begin{eqnarray}
\gamma_0\left[\frac{\partial \phi(x,t)}{\partial
x}\right]_{x=0} -\;\phi(0,t) + \delta\gamma(t)\left[\frac{\partial
\phi(x,t)}{\partial x}\right]_{x=0}=0\, . \label{bc-1} \nonumber \\ 
\end{eqnarray}
Also for the field, a perturbative approach will be adopted.
Following Ford and Vilenkin \cite{Ford-Vilenkin-1982} we
write
\begin{equation}
\phi(x,t)=\phi_{0}(x,t)+\delta\phi(x,t)\, , \label{ansatz}
\end{equation}
where, by assumption, $\phi_0$ satisfies the Klein-Gordon equation,
$\partial^2 \phi_0 = 0$, and the time-independent Robin BC,
\begin{equation}
\gamma_0\left[\frac{\partial \phi_0(x,t)}{\partial
x}\right]_{x=0} -\;\phi_0(0,t) = 0\, . \label{bc-PhiZero}
\end{equation}
The small perturbation  $\delta \phi$ takes into account the
contribution to the total field $\phi$ caused by the time-dependence
of the Robin parameter, described by the function $\delta\gamma(t)$.
Since both $\phi$ and $\phi_0$ satisfy the Klein-Gordon equation, so
does $\delta\phi$, namely, $\partial^2\delta\phi = 0$. The BC
satisfied by $\delta\phi$ is  obtained, up to first order terms, by
substituting (\ref{ansatz}) into Eq. (\ref{bc-1}), which leads to
\begin{eqnarray}
\gamma_0\left[\frac{\partial \delta \phi(x,t)}{\partial
x}\right]_{x=0}- \;\delta\phi(0,t) =
 - \delta\gamma(t)\left[\frac{\partial \phi_{0}(x,t)}{\partial
x}\right]_{x=0}\, , \label{cond-delta-phi} \nonumber \\
\end{eqnarray}
where Eq. (\ref{bc-PhiZero}) was used. Hereafter it will
be convenient to work in the Fourier domain, such that
\begin{widetext}
\begin{eqnarray}
\Phi(x,\omega) &=& \int dt \,\, \phi(x,t)\; e^{i\omega t}\; ;
 \;\;\;\;\;
 \Phi_0(x,\omega) = \int dt \,\, \phi_0(x,t)\; e^{i\omega t}\;
 ;\cr\cr
\delta\Phi(x,\omega) &=& \int dt \,\, \delta\phi(x,t)\; e^{i\omega
t}\; ;
 \;\;\;\;\;\;\;
\delta\Gamma(\omega) \;= \int dt \,\, \delta\gamma(t)\; e^{i\omega
t}\; . \label{fourier-trans}
\end{eqnarray}
\end{widetext}
It is worth emphasizing at this moment that, by assumption,
$\delta\gamma$ is a prescribed function of $t$, so that
$\delta\Gamma(\omega)$ is known, in principle. Since $\phi_0(x,t)$
is the solution with time-independent Robin BC, this field is
already known, and so does its Fourier transform, which is given by
(for the region $x > 0$),
\begin{eqnarray}
\Phi_{0}(x,\omega) &=&
\sqrt{\frac{4\pi}{|\omega|(1+\gamma_0^2\omega^2)}}\;\Bigl[\sin(\omega
x)
 + \gamma_0\omega\cos(\omega x)\Bigr]\nonumber \\ &\times &\Bigl[\Theta(\omega)a(\omega)
 - \Theta(-\omega)a^{\dagger}(-\omega)\Bigr], \label{field-exp}
\end{eqnarray}
where $\Theta(\omega)$ is the Heaviside step function. The operators
$a(\omega)$ and $a^{\dagger}(\omega)$ satisfy the usual bosonic commutation
relation $[a(\omega),a^{\dagger}(\omega^{\prime})]=2\pi\delta(\omega-\omega^{\prime})$.

In order to obtain $\Phi(x.\omega) = \Phi_0(x, \omega) +
\delta\Phi(x,\omega)$ we need to compute $\delta\Phi(x,\omega)$,
which satisfies the Helmholtz equation,
\begin{equation}\label{Helmholtz}
\Bigl(\partial^2_x \,
 + \, \omega^2\Bigr)\,\delta\Phi(x,\omega) = 0\; ,
\end{equation}
and is submitted to the BC below, obtained by Fourier transforming Eq.
(\ref{cond-delta-phi}),
\begin{align}
\gamma_{0}\left[ \frac{\partial \delta\Phi(x,\omega)}{\partial x}
\right]_{x=0} -\; \delta\Phi(0,\omega) = \nonumber \\ - \int\frac{d\omega^{\prime}}{2\pi}\left[
\frac{\partial\Phi_{0}(x,\omega^{\prime})}{\partial x}
\right]_{x=0}\delta\Gamma(\omega-\omega^{\prime}).
\label{CondCont-deltaPhi}
\end{align}
A further condition that must be imposed to the solution
 of Eq. (\ref{Helmholtz}) for $x>0$ is that it will lead to a solution for $\phi(x,t)$ that
 must travel to the right, since $\delta\phi(x,t)$ must describe a contribution
 coming from the mirror, and not going towards the mirror.
The desired solution can be written in terms of Green functions.
 Following the procedure given in \cite{Mintz-JPA-2006-2} it can be shown that the in and out
fields, denoted respectively as $\Phi_{\mbox{\footnotesize{in}}}$
and $\Phi_{\mbox{\footnotesize{out}}}$, are related to each other
according to
\begin{align}
\Phi_{\mbox{\footnotesize{out}}}(x,\omega) =
 \Phi_{\mbox{\footnotesize{in}}}(x,\omega) +
 \frac{1}{\gamma_{0}}\Bigl[ G_{\mbox{\footnotesize{R}}}^{\mbox{\footnotesize{ret}}}(0,x,\omega) \nonumber \\-
  G_{\mbox{\footnotesize{R}}}^{\mbox{\footnotesize{adv}}}(0,x,\omega)\Bigr]
 \times \left\{ \gamma_{0}\left[\frac{\partial \delta \Phi(x,\omega)}{\partial x}\right]_{x=0} -\;
 \delta\Phi(0,\omega) \right\}\; , \nonumber \\
\label{mintz-2006}
\end{align}
where $G_{\mbox{\footnotesize{R}}}^{\mbox{\footnotesize{ret}}}(0,x,\omega)$
($G_{\mbox{\footnotesize{R}}}^{\mbox{\footnotesize{adv}}}(0,x,\omega)$)
is the retarded (advanced) Robin Green function, satisfying the time-independent
Robin BC at $x=0$. These Green functions are given, respectively, by
\begin{equation}
G_{\mbox{\footnotesize{R}}}^{\mbox{\footnotesize{ret}}}(0,x,\omega)
 =  \left(\frac{\gamma_0}{1-i\gamma_0\omega}\right)e^{i\omega x},
\label{green-func-1}
\end{equation}
and
\begin{equation}
G_{\mbox{\footnotesize{R}}}^{\mbox{\footnotesize{adv}}}(0,x,\omega)
 = \left(\frac{\gamma_0}{1+i\gamma_0\omega}\right)e^{-i\omega x}.
\label{green-func-2}
\end{equation}
Inserting Eqs. (\ref{field-exp}) (appropriately relabeled as
$\Phi_{\mbox{\footnotesize{out}}}$ and
$\Phi_{\mbox{\footnotesize{in}}}$), (\ref{CondCont-deltaPhi}), (\ref{green-func-1}) and (\ref{green-func-2}) into Eq.
(\ref{mintz-2006}), we can readily obtain the Bogoliubov
transformation between $a_{\mbox{\footnotesize{out}}}$ and
$a_{\mbox{\footnotesize{in}}}$ and its hermitean conjugates:
\begin{align}
a_{\mbox{\footnotesize{out}}}(\omega)=
a_{\mbox{\footnotesize{in}}}(\omega)-2i\sqrt{\frac{\omega}{1+\gamma_0^2\omega^2}}
\int_{-\infty}^{+\infty}\,\frac{d\omega^{\prime}}{2\pi}
\sqrt{\frac{\omega^{\prime}}{1+\gamma_0^2{\omega^{\prime}}^2}} \nonumber \\
\times 
\Bigl[\Theta(\omega^{\prime})a_{\mbox{\footnotesize{in}}}(\omega^{\prime})
-\Theta(-\omega^{\prime})a^{\dagger}_{\mbox{\footnotesize{in}}}(-\omega^{\prime})
\Bigr] \,\delta\Gamma(\omega-\omega^{\prime}). \nonumber \\
\label{transform}
\end{align}
Noting that the  annihilation operator $a_{\mbox{\footnotesize{out}}}(\omega)$ is given in
terms of the  annihilation and creation operators $a_{\mbox{\footnotesize{in}}}(\omega)$
and $a^\dagger_{\mbox{\footnotesize{in}}}(\omega)$, respectively, we conclude that the
state $|0_{\mbox{\footnotesize{in}}} \rangle$ is not annihilated by
the $a_{\mbox{\footnotesize{out}}}(\omega)$ operators. Consequently, we can state that
particles were created from an initial vacuum state due only to the
time-dependence of $\delta\gamma(t)$ in the BC (\ref{bc-1}) imposed
on the field by the static mirror. In fact, for
$\delta\gamma(t)=0$ for all times, which corresponds to a static
mirror imposing the standard time-independent Robin BC on the field, we have 
$a_{\mbox{\footnotesize{out}}}(\omega)=a_{\mbox{\footnotesize{in}}}(\omega)$ and no
particles will be created, as expected. The particle creation effect
will be further investigated in the next sections, where we will
choose a specific time-dependent expression for $\gamma(t)$ in order
to compute explicitly the corresponding spectral distribution of the created particles as well as the respective particle creation rate.

\section{Spectral distribution of the created particles}
\label{Spectrum}

We start by writing the spectral distribution of the created
particles as
\begin{equation}
\frac{dN(\omega)}{d\omega} d\omega= \frac{1}{2\pi}\,\langle
0_{\mbox{\footnotesize{in}}}\vert\,
a^{\dagger}_{\mbox{\footnotesize{out}}}(\omega)
a_{\mbox{\footnotesize{out}}}(\omega) \, \vert 0_{\mbox{\footnotesize{in}}}
\rangle d\omega, \label{spec}
\end{equation}
where $dN(\omega)/d\omega$ is the number of created particles with
frequency between $\omega$ and $\omega + d\omega$ ($\omega \geq 0$)
per unit frequency. From the previous definition for
$dN(\omega)/d\omega$, it follows immediately that the total number
of created particles from $t =-\infty$ to $t=+\infty$ is given by
\begin{equation}\label{Numero-Energia}
N=\int_0^{\infty} \frac{dN(\omega)}{d\omega}\, d\omega\,.
\end{equation}

From Eq. (\ref{transform}) and its hermitian conjugate 
$a^{\dagger}_{\mbox{\footnotesize{out}}}$, it is straightforward to
show that
\begin{align}
\frac{dN(\omega)}{d\omega} = \frac{2}{\pi}\left( \frac{
\omega}{1+\gamma_0^2\omega^2} \right)
\int_{-\infty}^{\infty}\frac{d\omega^{\prime}}{2\pi}\frac{\omega^{\prime}}{1+\gamma_0^2{\omega^{\prime}}^2} \nonumber \\ \times {\left\vert
\delta\Gamma(\omega-\omega^{\prime})  \right\vert}^2
\Theta(\omega^{\prime}). \label{dndo}
\end{align}
In what follows we will obtain the spectral distribution for
a particular case of $\delta\Gamma (\omega)$. With this purpose in mind, let us consider the following expression for $\delta\gamma(t)$,
\begin{equation}
\delta\gamma(t)=\epsilon_{0} \cos(\omega_{0}t)\,
 e^{-|\, t\,|/T},
\label{DeltaGamma(t)1}
\end{equation}
with $\omega_{0} T \gg 1$. This choice of $\delta\gamma(t)$ may
simulate, for instance, the changing magnetic flux through a SQUID
fixed at the extreme of a unidimensional transmission line, as in
Ref. \cite{Johansson}, where a Robin-like BC arises
naturally from quantum network theory applied to the system under
consideration.

The expression of  $\delta\Gamma(\omega)$, obtained by Fourier
transforming Eq. (\ref{DeltaGamma(t)1}),  contains, in the limit of
$\omega_{0} T \gg 1$, two sharped peaks around $\omega= \pm \,
\omega_{0}$, which can be approximated by Dirac delta
functions, leading to the result
\begin{equation}
{\left\vert \delta\Gamma(\omega) \right\vert}^{2}  \approx
\frac{\pi}{2}\epsilon^{2}_{0}T\bigl[ \delta(\omega-\omega_{0}) +
\delta(\omega+\omega_{0})\bigr]\, .
\end{equation}
Substituting the above result into  Eq. (\ref{dndo}),  we finally
obtain the  desired spectral distribuition,
\begin{eqnarray}
\frac{dN(\omega)}{d\omega}
 &=& \left(\frac{\epsilon^{2}_{0}T}{2\pi}\right)
  \frac{\omega\,(\omega_{0}-\omega)}{(1+\gamma_{0}^2\omega^{2})\left[
1+\gamma_{0}^2(\omega_{0}-\omega)^2
\right]} \nonumber \\ &\times &\Theta(\omega_{0}-\omega),\label{spec-1}
\end{eqnarray}
for this particular situation.

\begin{figure}[!ht]
\begin{center}
\includegraphics[scale=0.6,angle=0]{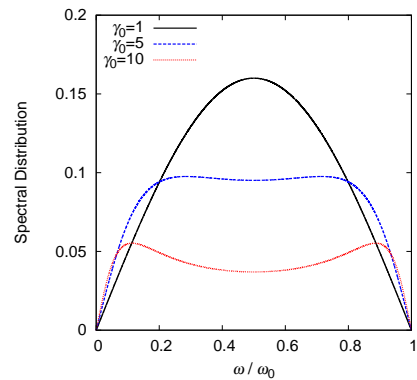}
\caption{ (Color online) The spectral distribution of the created particles $[(2\pi)/(\epsilon_{0}^{2} \, T)]\,dN/d\omega$ as a function of $\omega/\omega_{0}$ for several values of $\gamma_{0}$. Notice the reflexion symmetry around $\omega/\omega_{0}=0.5$: a signature of the fact that particles are created in pairs. The full line corresponds to $\gamma_{0}=1$; the dashed line to $\gamma_{0}=5$ and $20 \times [(2\pi)/(\epsilon_{0}^{2} \, T)]\,dN/d\omega$; and the dotted line to $\gamma_{0}=10$ and $100 \times [(2\pi)/(\epsilon_{0}^{2} \, T)]\,dN/d\omega$. }
\label{spc}
\end{center}
\end{figure}

A few comments are in order. Firstly,  observe (see Fig. \ref{spc} and Eq. (\ref{spec-1})) that ${dN(\omega)}/{d\omega}$ vanishes for
$\omega>\omega_{0}$, which means that no particles are created with
frequencies larger than $\omega_{0}$ $-$ the characteristic frequency
of the time-dependent BC. We also notice that the spectrum is left
invariant under the replacement $\omega \rightarrow
\omega_0 -\omega$. This is a signature of the fact that particles
are created in pairs: for each particle created with frequency
$\omega$ there is a twin particle created with frequency
$\omega_{0}-\omega$. In second place, note that for $\epsilon_{0} \rightarrow 0$, where a  Robin BC with a time-independent parameter $\gamma_0$ is re-obtained, the spectrum of created particles vanishes, as expected (recall that the mirror which imposes the BC on the field is at rest). Further, for a fixed (finite) value of $\omega_0$, the limit $\gamma_{0} \rightarrow \infty$ (Neumann BC imposed on the field at a static mirror) also leads to a vanishing spectrum of created particles. Finally, since we assumed $\epsilon_0\ll\gamma_0$, the limit $\gamma_{0} \rightarrow 0$ (Dirichlet BC imposed on the field by a static mirror) necessarily leads to a vanishing spectrum as well.

\section{Particle creation rate}
\label{creation_rate}

The total number of created particles is obtained by substituting
Eq. (\ref{spec-1}) in
(\ref{Numero-Energia}), namely,
\begin{eqnarray}\label{numerototal}
N &=& \left(\frac{\epsilon^{2}_{0}T}{2\pi}\right)
 \int_{0}^{\infty}
  \!\!\frac{\;\omega\,(\omega_{0}-\omega) \, \Theta(\omega_{0}-\omega)}{(1+\gamma_{0}^2\omega^{2})\left[
1 + \gamma_{0}^2(\omega_{0} - \omega)^2 \right]} \, \; d\omega \cr\cr
&=&
 \left(\frac{\epsilon^2_{0}\omega^{3}_{0}T}{2\pi}\right) F(\xi)\; ,
\end{eqnarray}
where $\xi=\gamma_{0}\omega_{0}$ and the function $F(\xi)$ is given by
\begin{equation}
F(\xi)=\frac{\left(2+\xi^2\right)\ln\left(1+\xi^2\right)-2\xi\arctan(\xi)}{\xi^{4}\left (4+\xi^2\right)}.
\end{equation}
As $N$ is proportional to $T$ - as expected for an open cavity - the physical meaningful quantity is the particle creation rate defined as $R=N/T$, that is 
\begin{equation}
R=\left(\frac{\epsilon^2_{0}\omega^{3}_{0}}{2\pi}\right) F(\xi)\;.
\label{rate}
\end{equation}
In the limits $\gamma_{0}\omega_0 \ll 1$ and $\gamma_{0}\omega_0 \gg 1$, the particle creation rate are approximately given by
\begin{equation}
R \approx \left(\frac{\epsilon^2_{0}\omega^{3}_{0}}{12\pi}\right) \;\;\;\; \mbox{for $\;\;\gamma_{0}\omega_0 \ll 1$}
\label{limites-1}
\end{equation}
\begin{equation}
R \approx \left(\frac{\epsilon^2_{0}\omega^{3}_{0}}{2\pi}\right)\frac{2\ln(\xi)}{\xi^4} \;\;\;\; \mbox{for $\;\;\gamma_{0}\omega_0 \gg 1$}.
\label{limites-2}
\end{equation}

For the sake of comparison with Eq. (\ref{rate}), we recall the total particle creation rates for moving mirrors with Dirichlet \cite{Jaekel-PRL-1996} (or equivalently for Neumann BC as proved in \cite{equivalencia-DN}})
\begin{equation}
R_{\mbox{\footnotesize{D/N}}}=\frac{\delta q^{2}_{0}\omega^{3}_0}{12\pi},
\label{lambrecht}
\end{equation}
and for time-independent Robin BC \cite{Mintz-JPA-2006-2} 
\begin{equation}
R_{\mbox{\footnotesize{ti-R}}}=\left( \frac{\delta q^{2}_0 \omega_0^3}{2\pi} \right)	G( \gamma_0\omega_0),
\label{mintz}
\end{equation}
where
\begin{align}
G(\xi)=\frac{\xi\left[  4\xi +\xi^3 + 12 \arctan(\xi)\right]-6\left( 2+\xi^2 \right)\ln\left( 1 + \xi^2 \right)}{6\xi^2 \left( 4 + \xi^2 \right)}. \nonumber \\
\end{align}
The formulas above were obtained assuming a non-relativistically small amplitude oscillatory law of motion for the mirror. For both cases $\delta q_0$ is the amplitude  and $\omega_0$ is the frequency of oscillation. We remark that for $\gamma_{0}\omega_0 \ll 1$ the particle creation rate in our model is exactly the same as for of a moving mirror \cite{Jaekel-PRL-1996} with Dirichlet BC  where $\epsilon_{0}$ plays the role of the amplitude of oscillation of the motion. This reinforces the possibility of simulating moving boundaries through a static mirror with time-dependent Robin BC. The three particle creation rates are compared in Fig. \ref{creationrate1}.

\begin{figure}[!t]
\begin{center}
\includegraphics[scale=0.6,angle=0]{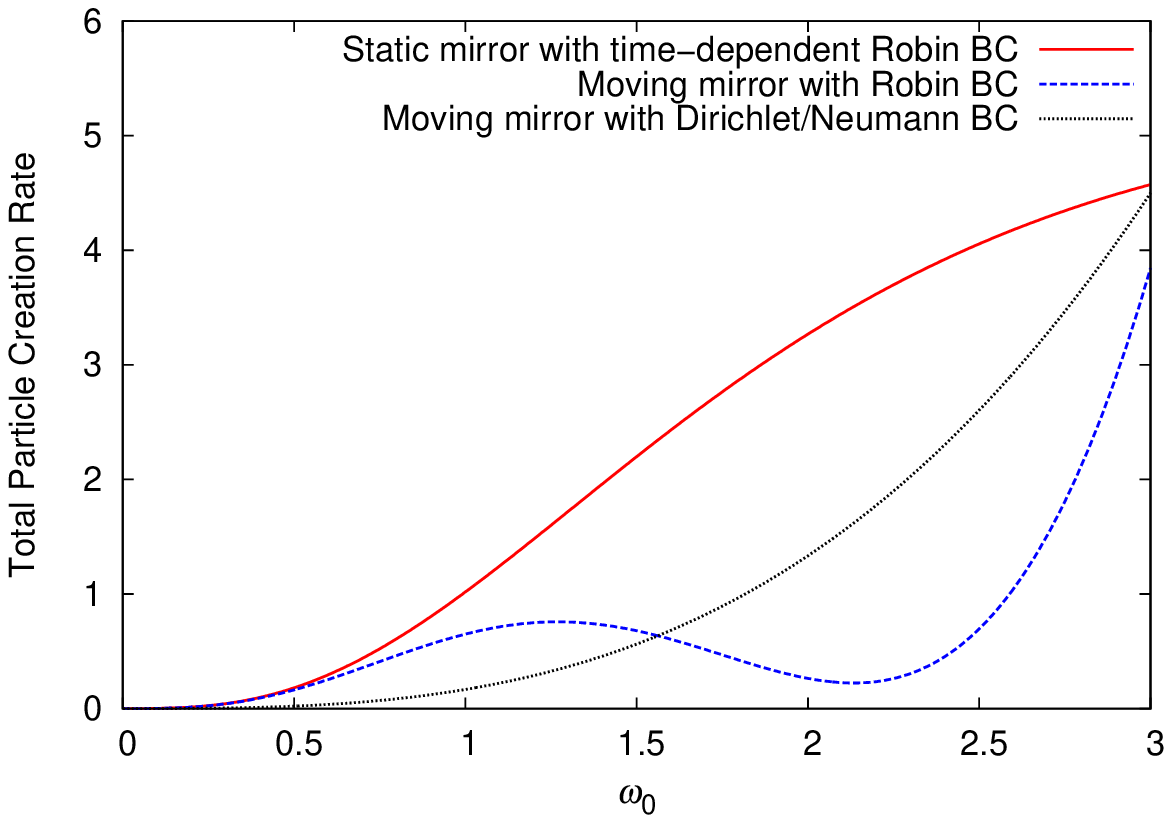}
\caption{(Color online) Comparison between the total particle creation rates given by Eqs. (\ref{rate}), (\ref{lambrecht}) and (\ref{mintz}). The full line corresponds to scaled creation rate $10\times \left[ \left( 2\pi\gamma_0^3 \right) / \left( \epsilon_0^2 \right) \right]\,R$. The dashed line corresponds to $10\times \left[ \left( 2\pi\gamma_0^3 \right) / \left( \delta q_0^2 \right) \right] \, R_{\mbox{\footnotesize{ti-R}}}$. Finally, the dotted line corresponds to $\left[\left( 2 \pi \right) / \left( \delta q_0^2 \right)\right]\,R_{\mbox{\footnotesize{D/N}}}$. In both curves involving Robin BC we considered $\gamma_0 =1$.}
\label{creationrate1}
\end{center}
\end{figure}

\begin{figure}[!t]
\begin{center}
\includegraphics[scale=0.6,angle=0]{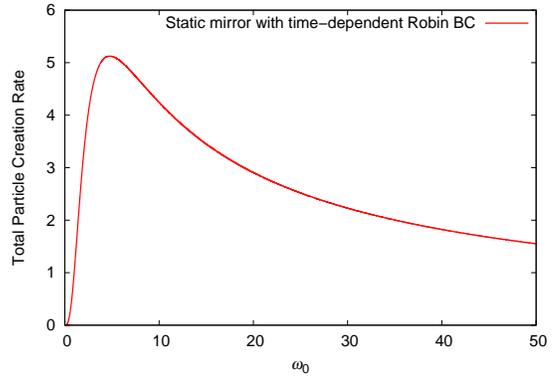}
\caption{(Color online) Behavior of Eq. (\ref{rate}) for a larger range of $\omega_0$, assuming $\gamma_0=1$. The creation rate approaches smoothly to zero as $\omega_0 \rightarrow \infty$. This does not happen for the other cases showed in the Fig. \ref{creationrate1}: for \textit{moving} mirrors the particle creation rate increases unlimitedly for larger values of $\omega_0$.}
\label{creationrate2}
\end{center}
\end{figure}

It is worth noting that the particle creation rate shown in Fig. \ref{creationrate2} starts growing with $\omega_0$ until it achieves a maximum value for a given value of $\omega_0$ and then it approaches monotonically to zero as $\omega_0$ goes to infinity. This behaviour should be compared with that obtained for a moving mirror which imposes on the field a Robin BC with a time-independent parameter, where the particle creation rate after passing through one maximum and one minimum grows indefinitely as $\omega_0$ goes to infinity (see Ref. \cite{Mintz-JPA-2006-2}). Naively, we could expect similar behaviors for these two problems, after all, a time-dependent Robin parameter should simulate, in principle, a moving mirror so that a high frequency oscillating $\gamma(t)$ should mean a high frequency oscillating mirror. However, the interpretation of the Robin parameter $\gamma$ as an estimative of the penetration depth of the material boundary is rigorously proved only for static mirrors. Even in this case, this identification is valid only for the field modes whose frequencies are much smaller than the plasma frequency (but this condition is easily achieved since the plasma frequency is much higher than the mechanical frequencies we want to simulate). It is plausible that such an interpretation remains valid for \textit{slowly} time-varying $\gamma (t)$, but not for high frequency oscillating $\gamma(t)$. In fact, our results show that this interpretation for $\gamma(t)$ fails for high values of $\omega_0$.

\section{Conclusions and final remarks}
\label{Conc}
 
Exploring the peculiar properties of Robin BC, particularly, the interpretation of the Robin parameter, we presented a simple and yet instructive theoretical model where a single static mirror with time-dependent properties described by a time-dependent Robin parameter simulates a moving boundary. We used this model to study analytically the dynamical Casimir effect of a system that may of some value for further understanding of a ongoing experiment based on a one-dimensional transmission line terminated by a SQUID. In this setup a time-dependent magnetic flux through the SQUID gives rise to particle creation phenomenon.
Employing  a perturbative approach, we showed that particles can be created due to the time-dependence of the Robin parameter $\gamma$. We obtained explicitly the spectrum of the created particles as well as the total particle creation rate for a particular choice of $\gamma(t)$ which has a practical interest concerning the experiment just described. Our model can also be used as a theoretical model to investigate other experimental setups suggested for measuring the dynamical Casimir effect, as for example, the promising experimental proposal of the Padua group \cite{Experimento-MIR}. All we have to do is to choose appropriately the time-dependence of $\gamma(t)$ to simulate correctly the physical situation under consideration.

We emphasize that the particle creation phenomena due to a time-dependent Robin BC imposed on the field at a static mirror has similarities and differences with the case where a time-independent Robin BC is imposed on the field at a moving mirror, as discussed by Mintz \textit{et al} \cite{Mintz-JPA-2006-2}. The main difference being the respective behaviors of the total particle creation rate for high values of $\omega_0$ (in the former case, where $\omega_0$ means the mechanical frequency of the moving mirror, this rate grows indefinitely as $\omega_0\rightarrow\infty$, while in the latter case, where $\omega_0$ gives a measure of how quick the time-dependent Robin parameter $\gamma(t)$ varies, this rate goes to zero, as $\omega_0\rightarrow\infty$). In the appropriate limits of the usual time-independent Dirichlet
($\gamma_{0} \rightarrow 0$), Neumann ($\gamma_{0} \rightarrow
\infty$) and Robin ($\delta\gamma(t)=0$) BCs no particles are
created, as expected. The generalization of the present work for 3+1
dimensions and cavities are also expected to have induced photon 
creation. These issues are under investigation  and
will be discussed elsewhere.

\section*{Acknowledgements}
The authors would like to thank the Brazilian agencies CNPq and Capes for a partial financial suport. H. O. S. would also like
to thank the hospitality of the Theoretical Physics Department of the Federal
University of Rio de Janeiro where part of this work was done. The authors are also grateful to A. L. C. Rego, D. T. Alves and T. Hartz for valuable discussions.


\end{document}